\documentclass[pdflatex,sn-mathphys-num]{sn-jnl}

\usepackage{graphicx}%
\usepackage{multirow}%
\usepackage{amsmath,amssymb,amsfonts}%
\usepackage{amsthm}%
\usepackage{mathrsfs}%
\usepackage[title]{appendix}%
\usepackage{xcolor}%
\usepackage{textcomp}%
\usepackage{manyfoot}%
\usepackage{booktabs}%
\usepackage{algorithm}%
\usepackage{algorithmicx}%
\usepackage{algpseudocode}%
\usepackage{listings}%
\usepackage{svg}
\usepackage{tabularx}
\usepackage{array}
\usepackage{booktabs}
\usepackage{multirow}
\usepackage{graphicx}
\usepackage[table]{xcolor}
\usepackage{rotating}
\usepackage{tikz}
\usepackage{helvet}
\usepackage{csquotes}
\usepackage{url}
\usepackage{hyperref}

\renewcommand{\orcidlogo}{%
  \raisebox{-0.14ex}{\includegraphics[height=1em,keepaspectratio]{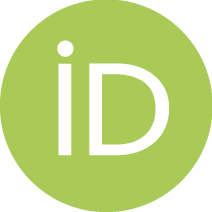}}%
}

\definecolor{agentfill}{HTML}{E1F5EE}
\definecolor{agentline}{HTML}{0F6E56}
\definecolor{agenttext}{HTML}{085041}
\definecolor{ctxfill}{HTML}{FAEEDA}
\definecolor{ctxline}{HTML}{BA7517}
\definecolor{ctxtext}{HTML}{633806}
\definecolor{hybfill}{HTML}{F1EFE8}
\definecolor{hybline}{HTML}{888780}
\definecolor{hybtext}{HTML}{444441}
\definecolor{bandbg}{HTML}{F7F6F2}
\definecolor{bandborder}{HTML}{D9D7CF}
\definecolor{textpri}{HTML}{2C2C2A}
\definecolor{textsec}{HTML}{5F5E5A}
\definecolor{tagcol}{HTML}{9C9A92}

\tikzset{
  band/.style={rounded corners=3pt, draw=bandborder, fill=bandbg, line width=0.4pt},
  lname/.style={anchor=west, font=\sffamily\small\bfseries, text=textpri},
  lsub/.style={anchor=west, font=\sffamily\scriptsize, text=textsec},
  ltag/.style={anchor=west, font=\sffamily\scriptsize, text=tagcol},
  agentchip/.style={draw=agentline, fill=agentfill, text=agenttext, rounded corners=2pt,
                    minimum height=0.58cm, font=\sffamily\footnotesize, inner sep=2pt,
                    anchor=west, line width=0.4pt, align=center},
  ctxchip/.style={draw=ctxline, fill=ctxfill, text=ctxtext, rounded corners=2pt,
                  minimum height=0.58cm, font=\sffamily\footnotesize, inner sep=2pt,
                  anchor=west, line width=0.4pt, align=center},
  hybchip/.style={draw=hybline, fill=hybfill, text=hybtext, rounded corners=2pt,
                  minimum height=0.58cm, font=\sffamily\footnotesize, inner sep=2pt,
                  anchor=west, line width=0.4pt, align=center},
  subagent/.style={draw=agentline, dashed, text=agenttext, rounded corners=2pt,
                   minimum height=0.58cm, font=\sffamily\footnotesize, inner sep=2pt,
                   anchor=west, line width=0.4pt, align=center},
  subhyb/.style={draw=hybline, dashed, text=hybtext, rounded corners=2pt,
                 minimum height=0.58cm, font=\sffamily\footnotesize, inner sep=2pt,
                 anchor=west, line width=0.4pt, align=center},
  legendsw/.style={minimum width=0.5cm, minimum height=0.32cm, inner sep=0,
                   anchor=west, line width=0.4pt},
  legendtx/.style={anchor=west, font=\sffamily\footnotesize, text=textsec},
}

\theoremstyle{thmstyleone}%
\theoremstyle{thmstyletwo}%

\theoremstyle{thmstylethree}%

\begin{document}

\title[Article Title]{A Technical Taxonomy of LLM Agent Communication Protocols}





\author*{\fnm{Linus} \sur{Sander} \orcid{https://orcid.org/0009-0007-4066-8194}} 
\email{linus.sander@tum.de}

\author*{\fnm{Habtom Kahsay} \sur{Gidey} \orcid{https://orcid.org/0000-0001-5802-2606}}
\email{habtom.gidey@tum.de}

\author{\fnm{Alexander} \sur{Lenz}}
\email{alex.lenz@tum.de}

\author{\fnm{Alois} \sur{Knoll} \orcid{https://orcid.org/0000-0003-4840-076X}}
\email{knoll@tum.de}

\affil{\orgname{Technische Universität München},
\orgaddress{\city{Munich}, \country{Germany}}}


\abstract{
As large language models (LLMs) advance and multi-agent systems aim to overcome the limits of standalone agents, robust communication protocols are becoming essential infrastructure for distributed agent networks. Nonetheless, the fragmented protocol landscape presents a significant interoperability challenge. This study develops a technical taxonomy to classify and analyze LLM agent communication protocols.
%
Following an established iterative method, we defined the taxonomy's purpose, meta-characteristic, and ending conditions, then performed five iterations, three empirical-to-conceptual and two conceptual-to-empirical, on nine actively maintained open-source protocols with demonstrable adoption.
%
The taxonomy comprises five dimensions: counterparty, payload, interaction state, discovery mechanism, and schema flexibility. Classification reveals recurring architectural patterns: all sampled agent-to-agent protocols combine hybrid payloads with session-state persistence; most protocols support multiple predefined schemas, and two negotiate schemas at runtime, indicating a trend toward schema flexibility; decentralized discovery remains rare.
%
Analysis suggests short-term convergence pressure toward protocols unifying agent-to-agent and agent-to-context (tool and data) communication. Long-term, however, no single protocol is likely to maximize versatility, efficiency, and portability simultaneously. The field will more likely evolve toward a federated, layered protocol stack. The framework guides protocol selection and highlights open research gaps such as privacy and policy enforcement.}



\maketitle
\section{Introduction}\label{sec:introduction}
Recently, large language model (LLM) based agents have attracted significant attention in AI research~\cite{han2024llm,masterman2024landscape,xi2025rise}. LLMs incorporate extensive world knowledge~\cite{yu2024kola} along with advanced reasoning and planning capabilities~\cite{achiam2023gpt,bang2023multitask,valmeekam2023planbench}. Embedded as the agent's core and equipped with memory, sensors, and actuators, the LLM grants the agent the capacity to interact dynamically with its environment and to perform efficient multistep reasoning, thereby enabling it to solve complex real-world tasks~\cite{liu2024agentbench,sun2023adaplanner,hsiao2025critical}.

Building on single LLM agents, a promising next frontier involves multiple cooperating agents. Such multi-agent systems (MAS) consist of specialized agents that communicate, collaborate, and debate with one another~\cite{guo2024large,talebirad2023multi}. Rooted in the principles of collective intelligence~\cite{malone2015handbook}, this approach further enhances the system's problem-solving capabilities, yielding outcomes superior to those achievable by any individual agent~\cite{guo2024large,chen2024agentverse,hong2024metagpt}. Here, one detail must not be overlooked: without a communication mechanism, no collective intelligence can arise. This crucial piece of infrastructure serves as the backbone that enables collective intelligence in MAS~\cite{guo2024large,han2024llm,yan2025beyond}.

To build and explore such systems, various frameworks exist that allow developers to construct MAS at varying levels of abstraction for diverse applications. Widely used examples include Microsoft's \textit{AutoGen}~\cite{autogengit,wu2024autogen}, \textit{CrewAI}~\cite{crewaigit}, CAMEL~\cite{camelgit,li2023camelcommunicativeagentsmind}, and \textit{LangGraph}~\cite{langgraphgit}. This variety of frameworks, however, has introduced a new challenge: a common protocol becoming necessary to standardize agent communication. Such standardization is a fundamental step toward unlocking the full potential of distributed MAS, where heterogeneous agents can dynamically discover each other and collaborate seamlessly across diverse use cases~\cite{cemri2025multi,yan2025beyond,kong2025survey}. 
Moreover, this standardization should extend to access to external systems, such as APIs or services, to enable seamless agent integration~\cite{ehtesham2025survey}. To overcome the antipattern of hard-coded communication pipelines, such a protocol must define the mechanics by which agents interact consistently, efficiently, and securely, with both other agents and external systems~\cite{marro2024scalable,li2025llm,kong2025survey,yan2025beyond}.

As the field of LLM-based agents is still relatively new and rapidly evolving, no universally accepted standard has yet emerged. Numerous solutions overlap in capability and are not interoperable with one another. To date, explicit communication protocols have not been widely studied, yet they are becoming an essential prerequisite for efficient cooperation and scalable MAS~\cite{yan2025beyond,li2025llm,kong2025survey}.

In this paper, we investigate the emerging landscape of communication protocols for LLM-based agents by developing a comprehensive taxonomy and analyzing nine concrete protocol implementations. Our taxonomy provides a coherent framework for understanding and classifying these protocols, as well as for tracking future developments. Finally, we present our key findings and conclude with a brief outlook on future developments.
\section{Background} \label{sec:background}
This section lays the groundwork for developing the taxonomy. It first examines LLM-based agents and MAS, before turning to the foundations of communication protocols that enable standardized agent coordination and interaction.

\subsection{LLM-Based Agents} \label{sec:llm-based-agents}

Back in 1998, Russell~\cite{russell1998learning} argued that~\enquote{AI is about the construction of intelligent agents}, noting that such agents are crucial for coping with dynamic, partially observable, and stochastic environments. An agent is generally understood as an autonomous entity that perceives its environment and takes actions to achieve specified goals~\cite{wooldridge1995intelligent,mele1995autonomous}. While early AI research paid limited attention to agents, interest grew significantly from the mid-1980s onward~\cite{jennings1998roadmap}. Common defining characteristics of agents exhibiting genuinely intelligent behavior include autonomy, reactivity, proactivity, social ability, and strong reasoning capabilities~\cite{nwana1996software,russell1998learning,mele1995autonomous}.

\subsubsection{LLM-Centric Control}

With the rise of large language models (LLMs), LLM-based agents have emerged as a new paradigm~\cite{xi2025rise}. Trained on massive text corpora, models such as GPT-4 (2023)~\cite{achiam2023gpt}, DeepSeek-V3 (2024)~\cite{liu2024deepseek}, or Claude Opus 4 (2025)~\cite{anthropic2025claude4systemcard} have demonstrated impressive language understanding and generation capabilities. Beyond language processing, these models exhibit a range of emergent capabilities, including human-like reasoning, planning, decision-making~\cite{bang2023multitask,liu2024agentbench,achiam2023gpt}, self-reflection~\cite{chen2024teaching,madaan2023self}, zero-shot learning~\cite{kojima2022large}, creativity~\cite{li2025review} and more~\cite{zhao2023survey}.

Unlike narrow models, LLMs enable agents to often require little or no task-specific training data. They can perform informed actions out-of-the-box using their internal knowledge~\cite{brown2020language,jin2023tab,ruan2023tptu}. In practice, an LLM agent can receive a high-level goal in text and autonomously plan and act to accomplish it, using the language model itself to guide multistep reasoning and planning~\cite{besta2024graph,yao2023react,yao2023tree}. It is even argued that LLMs pave a promising path toward general AI agents~\cite{xi2025rise} and that agents~\enquote{further push the boundary of LLMs towards AGI}~\cite{zhang2023one}, with self-improvement, autonomy, and continual learning as key components.

A typical workflow for an autonomous LLM-based agent unfolds as follows. The agent first receives a \textit{task} or high-level goal as text. It then begins to \textit{think} by generating natural language, creating subtasks, and devising a plan. Next, it \textit{acts} within its environment, executing the plan through appropriate tools. Finally, by \textit{observing} the environment and reviewing outcomes, the agent gathers \textit{feedback}, enabling it to re-plan, learn, and perform another iteration as needed~\cite{yao2023react,wang2024survey}.

\subsubsection{Fundamental Components and Concepts}

\begin{figure}
  \centering
  \includegraphics[width=1.0\textwidth]{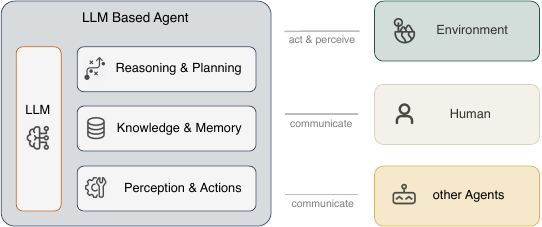}
  \caption{Fundamental components of LLM agents.} \label{fig-agent-system}
\end{figure}

One of the core ideas behind LLM agents is to combine language-based reasoning with structured control, and equipping the agent with components such as long-term memory and specialized tools~\cite{wang2024survey,masterman2024landscape,castrillo2025fundamentals}. 
Figure~\ref{fig-agent-system} provides a high-level overview of an agent system's key components and their connections to the external world.

\paragraph{Reasoning and Planning}

LLMs exhibit strong reasoning capabilities~\cite{valmeekam2023planbench}, which form the foundation of successful problem-solving, decision-making, and planning~\cite{evans1993reasoning,xi2025rise}. A wide range of techniques has been explored to enhance these capabilities further~\cite{huang2024understanding,masterman2024landscape}, ranging from task decomposition and planning approaches to continuous self-reflection loops that support adaptation. Task decomposition, in particular, helps an agent produce more granular and reliable plans for each subtask~\cite{wei2022chain,yao2023tree,besta2024graph}. 
Self-reflection mechanisms enable the agent to detect errors, revise its strategy, build resilience through adaptive self-healing, and tackle complex, long-horizon tasks~\cite{shinn2023reflexion,yao2023react,gidey2023modeling}.

\paragraph{Knowledge and Memory}

Thanks to the LLM acting as the central controller, the agent has access to a vast amount of encoded knowledge embedded within the foundation model's weights~\cite{petroni2019language,yu2024kola}. Beyond this, the agent also has access to in-context data, such as recent user conversations, records of tool invocations, and the agent's own thought history. This session-bounded data is commonly referred to as short-term memory or working memory~\cite{zhang2024survey}. Furthermore, the agent can be equipped with access to persistent external data sources, such as internal documents, external databases, and retrieval-augmented generation (RAG), sharing principles with document-based knowledge discovery~\cite{hatalis2023memory,lewis2020retrieval,zeng2024structural,gidey2022document}.

\paragraph{Perception and Action}

To operate effectively in a specific environment, an agent must be able to perceive that environment, recognize its affordances, and act within it~\cite{shen2024llm,gidey2025affordance}. Beyond the obvious input of textual information, multimodal perception can provide valuable additional capability. 
Visual input is particularly rich in information when the agent's operating sphere is a subset of the real world, necessitating resilient design patterns for visual processing~\cite{guo2022images,peng2023kosmos,pi2024perceptiongpt,gidey2026pattern}.

The interface to the environment is typically realized by providing tools the agent can use, marking a key difference between agentic systems and pure LLMs~\cite{ruan2023tptu,shen2024llm}. By generating output in a specific format, typically a JSON snippet, the agent can invoke a tool with the required parameters and receive the resulting output upon execution~\cite{shen2024llm}. Thanks to LLMs' powerful zero-shot and few-shot learning capabilities, new tools can be acquired simply by including their description in the prompt~\cite{ruan2023tptu,yao2023react}. Additionally, models fine-tuned for tool use have demonstrated strong effectiveness~\cite{shen2024llm,qin2023toolllmfacilitatinglargelanguage}.

\paragraph{Applications}

Equipped with these capabilities, LLM agents can operate far beyond simple conversational exchanges and traditional human-machine interactions~\cite{macedo2024evolving}. 
They can plan multistep processes, query knowledge sources, adapt over time, and more~\cite{huang2024understanding,zhang2024survey}. 
A variety of domains are now being tackled with specialized agents. These include computer control, ranging from code generation and terminal troubleshooting to GUI automation~\cite{bouzenia2025you,cao2024managing,qian2023chatdev,sager2025ai}. Knowledge work and cognitive automation, including innovation support, scientific workflow assistance, and knowledge exploration, are also now well within reach~\cite{li2025review,ghafarollahi2024sciagents,lala2023paperqare,katz2024knowledge,gidey2023user}. 
Across all of these areas, LLM agents are still at an early stage but are advancing rapidly~\cite{wang2024survey}.

\paragraph{Challenges}

Despite their rapid advancement, LLM-based agents still face significant limitations that must be addressed for robust real-world deployment. Fundamentally, they are not immune to the inherent vulnerabilities of underlying foundation models, such as hallucination~\cite{bang2023multitask,azamfirei2023large}, and finite context windows, which restrict the volume of historical data and environmental context they can process~\cite{hsieh2024ruler,liu2024lost,shi2023large}. Next, autonomously generated plans are not guaranteed to be feasible or efficient~\cite{huang2024understanding,kim2023language,xie2025revealing}, and the reliable integration of multimodal information sources remains an ongoing challenge~\cite{huang2024understanding}. Also, many LLM agent techniques heavily rely on pure prompting. Yet minor prompt variations can produce entirely different outputs ~\cite{chatterjee2024posix}.

Given the capabilities of LLM based agents and the abstract ideas of CI, the next natural step is to explore the collaborative potential of multiple agents~\cite{talebirad2023multi,tran2025multi}. For instance, one could query a smaller LLM N times to produce a set of independent answers. By selecting the most common answer, this method can, interestingly, outperform a single call to a state-of-the-art model~\cite{li2024more}. However, this captures only a narrow facet of CI, prompting a more fundamental question: what happens when we actively employ the communicative and social capacities of LLM agents to enable true cooperation? The following section addresses this question by a brief survey of the field of MAS.

\subsection{Multi-agent systems}
\label{sec:mas}

\subsubsection{LLM Based Multi-agent systems}

Multi-agent systems (MAS) consist of multiple agents that coordinate and communicate to accomplish a common goal. The idea is simple: independent, specialized agents jointly solve a task, thereby harnessing the benefits of collective intelligence~\cite{guo2024large,talebirad2023multi}. Decomposing complex objectives into manageable subtasks and assigning them to specialized agents naturally simplifies problem-solving~\cite{huang2024understanding,hong2024metagpt,khot2022decomposed}. Furthermore, diverse perspectives and varied interaction styles across agents, such as mutual feedback and debate, further enhance MAS problem-solving capabilities beyond what single-agent systems can achieve~\cite{tillmann2025literature,guo2026embodied,fu2023improving}. Such systems can tackle dynamic, complex tasks with strong performance and show emergent cognitive behavior~\cite{duenez2023social,guo2024large}.

\subsubsection{Key Concepts}

At a high level, an MAS is built from two essential elements: the agents themselves and the architecture that connects them. Each agent is characterized by a specific profile, including its role, abilities, and toolset, while the architecture serves as the core infrastructure that defines and enables collaboration among all agents~\cite{li2024survey,yan2025beyond}. Having outlined the core elements of individual LLM agents in Section~\ref{sec:llm-based-agents}, we now turn to the fundamental concepts of LLM-driven MAS.

The optimal topology of an agent ensemble remains a central research concern, bringing forth various architectural challenges in cognitive agent design~\cite{gidey2023towards}. 
Different \textit{architectures} exist, ranging from equi-level designs, in which all agents operate at the same hierarchical level, to hierarchical configurations, where one or several agents guide the group as a leader. 
Subsequently, the communication structure may be either fixed or dynamic, requiring rigorous methods to model such dynamic architectures~\cite{gidey2019modeling}. 
In fully dynamic systems, agent roles, relationships, and even the number of participating agents can evolve over time~\cite{han2024llm,masterman2024landscape,tillmann2025literature}.

Next, effective \textit{coordination and decision-making} is vital. Agents must debate, consult one another, align their goals, and resolve misunderstandings. Because MAS often yield multiple possible solutions and parallel discussions, techniques are required to converge on a single solution. Typical strategies include majority voting, consensus-seeking methods, or introducing a judge agent to make the final decision.~\cite{kaesberg2025voting,huang2024understanding}.

When looking more closely at the individual entities, the agents, \textit{profiling} becomes especially important in MAS. The specialization of each agent is critical for an effective system. Roles, characterizations, behavior descriptions, and skill sets must be carefully crafted, primarily through precise prompting and LLM model selection~\cite{wang2024survey,guo2026embodied,li2023camelcommunicativeagentsmind}. 

To successfully enable CI within MAS, a \textit{communication backbone} is crucial~\cite{guo2024large}. Without communication, agents cannot cooperate, debate, nor coordinate their actions. Ensuring that interactions are efficient and robust is therefore essential for achieving a reliable and high-performing MAS~\cite{han2024llm,yan2025beyond}. Therefore, given that MAS are expected to grow in size and distribution~\cite{wang2025internet,li2024more}, standardized communication protocols are critical to achieve scalability and flexibility without sacrificing robustness~\cite{ehtesham2025survey,kong2025survey}. They are key infrastructure elements that enable heterogeneous agents to connect reliably and dynamically. In essence, the communication infrastructure serves as the backbone enabling collective intelligence in MAS~\cite{marro2024scalable,yan2025beyond,guo2024large,cemri2025multi}.

\subsection{Communication Protocols}
\label{sec:communication-protocols}

In computer science, a \textit{protocol} is a formal set of rules or conventions that governs how entities, such as programs, processes, or agents, communicate and interact. It specifies the format, sequence, and meaning of the messages exchanged, enabling heterogeneous systems to understand one another straight away~\cite{pouzin2005tutorial,tanenbaum2011computer}. As Cerf and Kahn~\cite{cerf1974protocol} note, computers must~\enquote{share a common protocol (i.e., a set of agreed-upon conventions)} for data exchange to be meaningful.

In practice, a protocol specifies \textit{what} data is communicated, \textit{how} it is structured, and \textit{when} it should be sent or acknowledged. By defining these dimensions precisely, protocols enable independent, potentially heterogeneous systems to communicate reliably. Often, they also include procedures for error handling, synchronization, and more, so that, even under adverse conditions, reliable communication is ensured~\cite{pouzin2005tutorial,tanenbaum2011computer}.

The internet itself is the most obvious example of a protocol's importance. Foundational standards such as the \textit{Internet Protocol} (IP) and the \textit{Transmission Control Protocol} (TCP), together with the \textit{Hypertext Transfer Protocol} (HTTP), form today's backbone of global data exchange~\cite{nath2015tcp,cerf1974protocol}.

Beyond this, MAS require application-level protocols clarifying how autonomous agents interact. Before the rise of LLMs, well-known examples included the protocols \textit{Knowledge Query and Manipulation Language} (KQML)~\cite{finin1994kqml} and \textit{FIPA Agent Communication Language} (ACL)~\cite{labrou1999agent}. KQML, developed in the early 1990s, was introduced as~\enquote{a new language and protocol for exchanging information and knowledge} among software agents. Central to KQML is an extensible set of operations called \textit{performatives}, such as \textit{ask} and \textit{tell}, which lays the groundwork for higher-level interaction patterns like negotiation~\cite{finin1994kqml,labrou1999agent}. Its successor, ACL, represented the first organized effort to standardize communication in the broader agent community. Building on speech-act theory, ACL lets agents express intentions, requests, information, and proposals while each message follows a formal structure with fields for sender, receiver, content, and more, enabling interoperability across heterogeneous agents~\cite{labrou1999agent}.

In LLM-based MAS, standardized communication protocols become essential for maintaining consistent, secure, and reliable interactions among heterogeneous agents~\cite{yan2025beyond,ehtesham2025survey,kong2025survey}. As systems scale, these protocols become a prerequisite for robust multi-agent collaboration, enabling greater efficiency, adaptability, and overall performance with minimal human intervention~\cite{li2025llm,cemri2025multi}.

\section{Related Work} \label{sec:related-work}

Several recent works have examined the emerging landscape of LLM agent communication protocols, highlighting both the rapid progress and the significant challenges that remain. 

From a standardization standpoint, Li et al.~\cite{li2025llm} argue that the current fragmentation of agent communication resembles the early~\enquote{protocol wars} of networking and advocate for a unified framework. Next, Du~\cite{du2025ai} examines AI agent communication through the perspective of Internet architecture, distilling five key design principles, to guide the sustainable development of multi-agent ecosystems. Moreover, Kong et al.~\cite{kong2025survey} provide a comprehensive security-focused survey of agent communication, proposing a three-class taxonomy for categorizing agent communication, while systematically analyzing their associated vulnerabilities and potential defense mechanisms. Complementing those perspectives, Ehtesham et al.~\cite{ehtesham2025survey} offer a comparative analysis of four concrete interoperability protocols: MCP, ACP, A2A, and ANP, evaluating them across dimensions such as interaction modes, discovery mechanisms, and security models. 

Further research exists on concrete protocols. Among the sample that is later examined, the Agora protocol~\cite{agoragit} stands out for its scientific rigor and for emphasizing fundamental principles needed in an efficient, scalable, and robust design. In particular, Marro et al.~\cite{marro2024scalable} identified a \textit{trilemma} of three quality attributes, none of which can be fully optimized simultaneously. First, \textit{versatility} requires the protocol to support multiple message types, from text only to structured data. Second, \textit{efficiency} demands minimal computational and networking costs. Third, \textit{portability} is of focus, reducing the effort required for any arbitrary agent to adopt and use the protocol~\cite{marro2024scalable}.

Last but not least, Yang et al.~\cite{yang2025survey} also propose a taxonomy for LLM agent communication protocols. Their taxonomy comprises two dimensions: \textit{object orientation}, which distinguishes \textit{context-oriented} from \textit{inter-agent} protocols according to the intended counterpart, and \textit{application scenario}, which classifies protocols as either \textit{general-purpose} or \textit{domain-specific}. In addition to that, they collected a sample of 14 protocols and evaluated them against quality attributes such as efficiency, security, and scalability, identifying several emerging trends in agent development~\cite{yang2025survey}. Nevertheless, a taxonomy with just two dimensions seems insufficient to provide the abstract, hierarchical structure needed to explore, analyze, and understand the field in depth.

\section{Approach}\label{sec:study-approach}
\subsection{Taxonomy Development Method}\label{sec:methodology-taxonomy}

To develop the taxonomy itself, we followed the widely used method for taxonomy construction from Nickerson et al.~\cite{nickerson2013method}. 
This method was selected because it provides a structured, iterative, traceable, and reproducible procedure for constructing taxonomies, rather than relying on an ad hoc classification process~\cite{oberlander2019taxonomy}. 
We first begin this section by outlining and clarifying the qualities that define a useful taxonomy, then discuss the iterative procedure that drives its development as applied in this study. %

The term \textit{taxonomy} can be used in several ways. 
In this study, the term \textit{taxonomy} refers to a classification framework. Fundamentally, the goal is to provide a useful hierarchical classification system that organizes key concepts and their concrete values. It aims to capture the essence of a specific domain, thereby helping its users to understand, discuss, analyze, and observe complex concepts~\cite{nickerson2013method,oberlander2019taxonomy}. 
In contrast to a \textit{typology}, which is a conceptual classification approach, a taxonomy is empirical, derived from real-world data and observation~\cite{bailey1994typologies}. Following Nickerson et al.~\cite{nickerson2013method}, a taxonomy consists of several distinct dimensions, each of which contains two or more characteristics that should be mutually exclusive and collectively exhaustive~\cite{nickerson2013method,oberlander2019taxonomy}.

Nickerson et al.~\cite{nickerson2013method} define five qualitative attributes of a useful taxonomy.
A taxonomy should be 
\textit{concise}, enabling researchers to recall the entire taxonomy with ease; 
\textit{robust}, containing enough dimensions and characteristics to differentiate objects in the domain meaningfully; 
\textit{comprehensive}, able to classify every instance in the domain and contain all relevant object dimensions; 
\textit{extendible}, allowing new characteristics and dimensions to be added without difficulty when necessary; and 
\textit{explanatory}, offering dimensions and characteristics that provide a useful and explainable abstraction~\cite{nickerson2013method}.
These criteria served as subjective ending conditions for our taxonomy development process. 

In this study, we rigorously follow the clear development process to ensure quality and reproducibility. 
The entire development process began with the definition of three foundational elements. 
First, the researcher should think of the overall \textit{purpose} of the taxonomy: for what should it especially be used later and who are the main users. 
Next, a so-called \textit{meta-characteristic} has to be set up, to provide a basis for all characteristics to be constructed later. 
A typical wording could be: 'Classify … based on …' and can encapsulate multiple facets. 
After establishing this anchor, the researcher selects objective \textit{ending conditions} (e.g. no new dimensions emerge and every characteristic is instantiated) and subjective ending conditions (namely, the five quality criteria outlined above)~\cite{nickerson2013method}. 

After these elements had been defined, we followed Nickerson et al.'s iterative development cycle. 
The development cycle goes over, right in the beginning, to a fundamental decision where each iteration requires a choice between two development paths. 
In the \textit{conceptual-to-empirical} path, the researcher first identifies candidate dimensions from theory, applies them to a sample of concrete objects of the domain, and prunes or augments the constructed dimensions. 
On the alternative \textit{empirical-to-conceptual} path, the researcher instead examines a sample of concrete objects, distills meaningful attributes and properties, and groups these into abstract dimensions that satisfy the basic quality criteria of sound taxonomy design. 
Each iteration ends with a review and evaluation of the resulting taxonomy against the predefined ending conditions.
If these conditions are not satisfied, another iterative cycle begins, starting with the binary path choice~\cite{nickerson2013method}.

Conceptually, this process mirrors the build-evaluate cycle in design science research~\cite{hevner2004design}. 
Each iteration constructs a provisional taxonomy or schema and immediately evaluates it against empirical reality and the predefined quality criteria.
This allows usefulness, purpose-fit, and traceability to be integrated into the taxonomy development process. 
The approach also encourages detailed documentation of the entire procedure, which supports both reproducibility and extensibility. 
In our study, this disciplined iterative procedure, bounded by explicit rules, was used to yield a high-quality taxonomy with unique dimensions and mutually exclusive, collectively exhaustive characteristics~\cite{nickerson2013method,oberlander2019taxonomy}.

\subsection{Communication Protocols Sample}\label{sec:methodology-sample}

This section outlines how we selected the nine protocols that ground our taxonomy development and offers concise summaries of each.

First, it is important to define the general type and domain of the target protocols. Although we further clarify this when specifying the \textit{meta-characteristic} in the taxonomy development process, it is already essential for the protocol selection. Specifically, we searched for protocols that were explicitly developed to connect an LLM agent with another system. Consequently, more generic agent communication implementations and concepts, such as the \textit{FIPA Agent Communication Language}~\cite{o1998fipa} which implements speech-act theory, or the \textit{Contract Net Protocol}~\cite{smith1980contract}, an abstract task-sharing protocol for MAS, were excluded from our sample. Likewise, protocols not explicitly designed for LLM agents fall outside the scope of the taxonomy.

Next, we focus exclusively on protocols that are open source and have an implementation ready to be used. The implementation does not have to be production ready, a test or research prototype is sufficient. For instance, the Agora protocol~\cite{marro2024scalable} is not production ready, yet an implementation exists, whereas the LOKA protocol~\cite{ranjan2025loka} is just a proposed concept with no implementation. Therefore, the latter has been excluded. Also, Firecrawl~\cite{firecrawlonline} and the uAgents protocol~\cite{uagentsonline} are proprietary solutions and are not entirely open source and thus omitted. Emphasizing the open source requirement reflects the historical pattern that for major web standards it has been critical to be open source, in order to achieve widespread adoption and become a global standard~\cite{gamalielsson2024open}.

All selected protocols maintain publicly accessible codebases on GitHub, enabling verification of active maintenance status. GitHub stars are employed as a quantitative proxy for community adoption and real-world traction. For that reason, protocols such as the \textit{Agent Protocol}~\cite{agentprotocolgit}, whose last commit not just changing the README file occurred on Jun 11 2024 (accessed 7 Jul 2025), are treated as inactive and therefore excluded. Likewise, repositories such as the AITP protocol~\cite{nearaitpgit} are omitted, as they have accumulated only 20 \textit{GitHub stars} (accessed 7 Jul 2025) since their initial release in February 2025.

Nine protocols met all selection criteria and are outlined in the following section. Table~\ref{tab:protocol-classification-markers} and~\ref{tab:protocol-classification-text} provide a quick overview, and further characteristics are discussed during taxonomy development.

\paragraph{Model Context Protocol (MCP)}

Developed by Anthropic, the \textit{Model Context Protocol} (MCP) provides a standardized way to augment any LLM with tools and contextual information by enabling existing applications, APIs, and raw data sources, in short almost any IT component, to present their context and capabilities to LLMs in a standardized manner~\cite{mcpgit,mcpdocscorearchitecture}.

At its core, MCP follows a straightforward client-server model. Host applications, such as Claude Desktop or IDE plug-ins, embed \textit{MCP clients} that communicate with \textit{MCP servers}, each of which exposes specific capabilities, including data retrieval, tool invocation, and prompt delivery for the LLM behind the MCP client~\cite{mcpdocscorearchitecture}.

\paragraph{Agent to Agent (A2A)}

Google developed this protocol to address the rising number of siloed agent infrastructures. As agents are implemented and deployed across diverse frameworks and platforms, a standardized protocol is required to facilitate collaboration~\cite{a2adocskeyconcepts}. Its primary goal is to enable agent-to-agent connectivity, not to replace the already well-known MCP protocol~\cite{a2adocsmcp}.

Designed to be secure and extensible by default, it implements a variety of concepts across multiple levels of communication, supporting both immediate and extended tasks, including streaming updates, state synchronization, and artifact exchange (e.g. files and structured data)~\cite{a2agit,a2adocsspecification}. 

\paragraph{LangChain Agent Protocol (LAP)}

LangChain created a standardized API for deploying LLM agents in production. Specifically, they defined a RESTful API with a unified set of endpoints. The main endpoints are \textit{/runs} for executing agents, \textit{/threads} for managing multi-turn conversations, and \textit{/store} for long-term memory storage. Through these endpoints, any client can invoke, monitor, and persist agent workflows, irrespective of the agents’ internal implementations~\cite{lapgit,lapdocsapi}.

\paragraph{agents.json}

This simple protocol defines a JSON specification through which existing websites and APIs can be discovered and interpreted by LLM based agents. Built on top of the \textit{OpenAPI} standard, agents.json offers a structured, LLM-friendly, and stateless way to define and consume API workflows, letting agents execute multi-step tasks reliably with minimal boilerplate and without extensive prompt engineering~\cite{agentsjsongit,agentsjsonintro,agentsjsonschema}.

\paragraph{Agora}

Oxford researchers introduced the \textit{Agora Protocol}~\cite{agoragit} to enable decentralized collaboration among diverse LLM agents, without central servers and pre-defined communication schemas. Its design tackles the \textit{Agent Communication Trilemma}, balancing versatility (handling diverse message types), efficiency (minimizing token and API cost), and portability (working across agent platforms), in a single decentralized framework~\cite{marro2024scalable}. For instance, it includes an on-the-fly protocol negotiation mechanism. Before exchanging actual data, agents share or even generate new communication schemas~\cite{marro2024scalable,agoraspecification}.

\paragraph{Agent Network Protocol (ANP)}

Its vision is to become~\enquote{the HTTP of the agentic web era}~\cite{anpgit}. It is a peer-to-peer communication protocol designed to enable secure, decentralized collaboration among heterogeneous LLM-powered agents across the internet. The protocol is intentionally flexible and feature-rich out of the box, offering three layers: an \textit{Identity Layer} for secure communication, covering authentication and end-to-end encryption; a \textit{Meta Protocol Layer} that allows agents to negotiate communication styles dynamically, and an \textit{Application Layer} that supports straightforward agent interaction through standardized descriptions and management~\cite{anpcommunication,anptechwhitepaper}.

\paragraph{LMOS}

This protocol is part of the \textit{Eclipse LMOS} ecosystem and enables agents and tools to easily publish, connect, and share their capabilities. As a special feature, it standardizes metadata and interaction patterns while remaining agnostic to the underlying transport layer, allowing seamless operation over HTTP, WebSocket, MQTT, and other protocols. Moreover, it emphasizes that it can readily support an internet-scale MAS with thousands of tools and agents, an \textit{Internet of Agents}~\cite{lmosgit,lmosdiscovery}.

\paragraph{Agent Communication Protocol (ACP)}

The ACP is a RESTful open standard, created by BeeAI and IBM under Linux Foundation governance, that enables seamless, structured communication, discovery, and coordination among heterogeneous LLM agents~\cite{acpgit,acpcoreconcepts}.

ACP defines consistent HTTP/JSON interfaces that enable agent discovery, execution, and multimodal message exchange. It also implements so-called \textit{ACP Servers}, where agents can register themselves for discovery and can likewise discover other publicly available agents~\cite{acpcoreconcepts}.

\paragraph{agntcy}

Its mission extends beyond defining a protocol for agent-to-agent communication, it seeks to establish an open-source infrastructure for the \textit{Internet of Agents}~\cite{agntcygit}. For communication, it specifies the \textit{Agent Connect Protocol}, an API extending \textit{OpenAPI} that enables agents to connect, share state, stream results, and handle authentication and authorization seamlessly across frameworks and runtime environments. The potential standard also provides essential infrastructure elements, including a centralized discovery service, a fixed agent manifest format, and many more~\cite{agntcyconnectprotocol,agntcyintroduction,agntcydiscovery}.

\section{Results: Taxonomy Design} \label{sec:taxonomy-design}

We now proceed to construct the taxonomy by first defining its fundamental artifacts and summarizing the iterative development process. Rather than detailing each iteration, we then focus on examining both accepted and rejected dimensions in detail. Finally, we classify concrete protocols from our sample to illustrate the taxonomy's dimensions and their application.
\subsection{Developing the Taxonomy}
\label{sec:taxonomy-design-developing-the-taxonomy}

\subsubsection{Scope} 

As discussed in Section~\ref{sec:study-approach}, Nickerson et al.'s technique~\cite{nickerson2013method} begins with the definition of three foundational artifacts that define the scope of the taxonomy.

The \textit{purpose} of the taxonomy is to classify and distinguish LLM agent communication protocols based on the type of communication they facilitate. The taxonomy focuses on protocols that connect an LLM agent to another system, such as another agent or a general information system. It operates primarily at the application level rather than the solution level, and therefore does not delve into concrete technical implementation details.

The \textit{expected users} of the taxonomy are researchers and agent developers, as well as non-experts who simply want to access an agent through an API. Potential users may wish to obtain an overview of the forms of communication and guidance for protocol selection. In addition, a brief overview of how protocols differ overall, and where gaps and opportunities for further research and development exist, is of interest.

The \textit{meta-characteristic} is defined as classifying the protocols according to the types of components involved and the characteristics of their communication.

\subsubsection{Process and Quality}

In total, we conducted five iterations: the first three followed the empirical-to-conceptual approach, and the final two followed the conceptual-to-empirical approach. Each of the first three iterations examined three of the nine collected protocols. The final two iterations complemented the analysis by consolidating accumulated knowledge and testing more abstract dimensions.

Ultimately, the iterative approach, guided by clear ending conditions and the taxonomy's purpose and meta-characteristic, made development straightforward and appeared to strengthen its robustness and comprehensibility.

With respect to the basic quality criteria of a qualitative taxonomy, we focused on the following key properties. The developed taxonomy is concise, as it contains a clear and manageable number of dimensions and characteristics. 
It is also robust, in the sense that we examined exactly nine protocols in detail and discovered no new valid dimensions in the final iteration. Furthermore, comprehensiveness and explanatory quality are ensured, as we demonstrate that all protocols can be classified according to the taxonomy in a straightforward manner. Finally, the taxonomy satisfies the foundational principle of mutual exclusivity and collective exhaustiveness, which we systematically enforced during dimension construction.


\subsection{Accepted Dimensions}
\label{sec:taxonomy-design-accepted-dimensions}

Throughout the iterative process, various dimensions were examined based on discovered common and distinctive characteristics. 
All considered dimensions are now described in detail, separated into accepted and rejected ones.

For each dimension, we present its corresponding values and assess whether it satisfies mutual exclusivity and exhaustiveness. For accepted dimensions, we additionally cite foundational research supporting the dimension and its values.

\begin{figure} 
  \centering
  \includegraphics[width=1.0\textwidth]{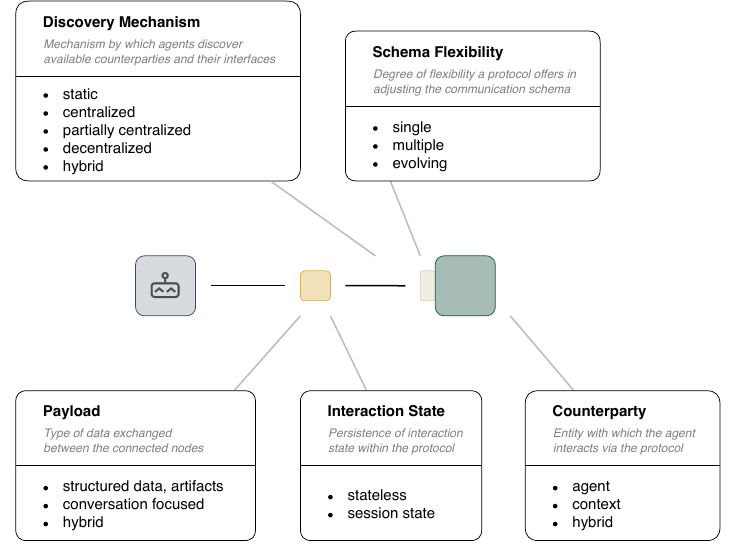}
  \caption{Finalized Taxonomy: all dimensions and characteristics} \label{fig-taxonomy}
\end{figure}

The finalized taxonomy comprises five dimensions, as shown in Figure~\ref{fig-taxonomy}.

\subsubsection{Counterparty}

We begin with the most self-explanatory dimension: \textit{counterparty}. It identifies the type of entity with which the agent interacts via the protocol. Its values are: \textit{agent}, where the protocol connects one LLM agent to another; \textit{context}, where non-agent entities such as tools, services, APIs, or data sources are connected; and \textit{hybrid}, where both are supported.

The dimension is mutually exclusive and exhaustive, as \textit{context} captures any non-agent counterparty by design. Although further subcategories within \textit{context} could be defined, the dimension is intentionally kept concise.

This dimension aligns with Yang et al.'s~\cite{yang2025survey} protocol taxonomy. Their \textit{object-orientation} dimension distinguishes \textit{context-oriented} from \textit{inter-agent} protocols, corresponding to our \textit{context} and \textit{agent} values, respectively. We extend their scheme with a third value, \textit{hybrid}, to accommodate protocols supporting both types.

\subsubsection{Payload}

Next, the \textit{payload} dimension classifies the kind of data a protocol exchanges. A protocol falls into one of three categories: \textit{structured data and artifacts}, where only structured data or artifacts, including meta-information, are exchanged; \textit{conversation focused}, where the payload is message-centric and text is always part of the exchange; or \textit{hybrid}, where both are supported.

To clarify this dimension further, we compare it with the three basic data types: structured, semi-structured, and unstructured~\cite{eberendu2016unstructured}. Rather than adopting that typology, our taxonomy classifies payloads according to the intended purpose a protocol is designed to support. A protocol is labeled \textit{conversation focused} when text is always part of the payload and structured or semi-structured data appear only as optional extensions. Conversely, if the protocol can solely transmit structured data and artifacts drawn from any of Eberendu's~\cite{eberendu2016unstructured} three data types, it is assigned to the \textit{structured data and artifacts} category.

\subsubsection{Interaction State}

This dimension captures the persistence of \textit{interaction state} within a protocol, that is, whether the protocol implements a stateful unit of work between connected components. A protocol is either \textit{stateless} or provides \textit{session state} functionality.

Mutual exclusivity and exhaustiveness are guaranteed because the dimension is binary. A protocol that preserves context across messages is classified as \textit{session state}, one that lacks such mechanisms is \textit{stateless}.

Ling et al.~\cite{ling2004session} identify four kinds of managed state in communication networks: stateless, soft state, session state, and hard state. Their definition of stateless, no information retained at the nodes, matches our equivalently named value directly. Session state, defined as information persisting only for the duration of a user session, corresponds to what we term \textit{session statefulness}~\cite{ling2004session}. The remaining categories, hard state, referring to long-term memory, and soft state, denoting temporarily cached information, fall outside the scope of this dimension. The dimension only concerns the duration of a single session and not what persists beyond it. This is further discussed in the context of persistent state in Section~\ref{sec:taxonomy-design-rejected-dimensions}.

\subsubsection{Discovery Mechanism}

In dynamic, large-scale distributed systems, individual nodes are typically unaware of the precise identities or locations of other nodes. Therefore, efficient discovery mechanisms are essential for scalable and flexible agent networks~\cite{ahmed2010survey,yang2025survey}.

We identify two initial values: \textit{static}, where the requesting agent must know the endpoint a priori, and \textit{centralized}, where a registry maintains a list of accessible endpoints.

To ensure exhaustiveness, we drew on Ahmed and Boutaba's~\cite{ahmed2010survey} survey of distributed search techniques in large-scale systems. They index two centralized content-sharing architectures, registry and index, which we combine into a single value, \textit{centralized}, since we only are concerned with the discovery mechanism itself, not with how the registry is built. Their classification further motivates two additional values: \textit{partially centralized}, where a limited set of supernodes assists with discovery, and \textit{decentralized}, where no central authority exists and each peer maintains its own index, with discovery occurring via network broadcasting. Finally, we add a \textit{hybrid} value for protocols that support multiple discovery mechanisms beyond static configuration.

Mutual exclusivity is guaranteed by arranging the values hierarchically. \textit{Static} represents the baseline. If a protocol offers a \textit{centralized}, \textit{partially centralized}, or fully \textit{decentralized} discovery mechanism, it is classified as such, even though static configuration remains universally available. A protocol supporting multiple non-static mechanisms is assigned to \textit{hybrid}.

\subsubsection{Schema Flexibility}

Next, we introduce the \textit{schema flexibility} dimension, which captures the degree of flexibility a protocol offers in adjusting communication schemas, message formats, or dialogue structure. A protocol may allow only a \textit{single} schema, one basic interaction pattern defined before runtime. Alternatively, it may support \textit{multiple} static schemas, established a priori, from which the requester can choose at runtime. Finally, if the interacting components can negotiate new interaction styles at runtime, the protocol is classified as \textit{evolving}.

In this dimension, the term \textit{schema} follows the standard computer-science definition: a formal structure representing an engineered artifact~\cite{bernstein2011generic}. In our context, it refers in particular to the structure of the exchanged content. For instance, a tool invoked by an agent may require a fixed schema with a predefined set of typed variables~\cite{mcpdocscorearchitecture}, whereas an LLM-based agent might accept only a single text input, subject to constraints such as a maximum length~\cite{lapdocsapi}.

It is important to note that this dimension does not characterize the structural nature of the payload, that role belongs to the \textit{payload} dimension. Instead, it specifies whether exchanged information must conform to a fixed schema, whether agents may select from multiple predefined schemas, or whether the protocol supports runtime negotiation of schema structure.


\subsection{Rejected Dimensions}
\label{sec:taxonomy-design-rejected-dimensions}

Having examined all accepted dimensions, we now turn to three rejected dimensions and explain the rationale for their exclusion from the taxonomy.

\subsubsection{Initiative Flow}

This dimension captures which party may initiate communication. Its proposed values are: \textit{unidirectional}, where only the client can initiate; \textit{context-bound bidirectional}, where either party may send messages within an active session but only one side can initiate it; and \textit{open bidirectional}, where any party can initiate communication at any time.

The dimension is rejected on two grounds: most protocols fall under \textit{context-bound bidirectional}, and any protocol could in principle be implemented symmetrically to support open bidirectional initiation. It therefore fails the test of mutual exclusivity and adds little analytical value.

\subsubsection{Simplicity} 

The examined protocols differ notably in their simplicity and intended lightweightness. Some, such as \textit{A2A} and \textit{LMOS}, incorporate a broad range of functionalities~\cite{a2adocsspecification,lmosintroduction}, whereas others, such as \textit{Agora}, \textit{agents.json}, and \textit{ACP}, target specifically defined use cases and explicitly prioritize a lightweight design~\cite{agoraspecification,agentsjsonintro,acpcoreconcepts}.

Nevertheless, this dimension poses an immediate problem: there are no clear boundaries to ensure mutually exclusive values. The notion of simplicity is too vague to construct a meaningful qualitative dimension, and it is therefore rejected.

\subsubsection{Persistent State}

This dimension classifies protocols by their ability to store data about prior interactions beyond a single or multi-turn exchange. Unlike the \textit{interaction state} dimension (Section~\ref{sec:taxonomy-design-accepted-dimensions}), it concerns exclusively long-term statefulness. The stored content might include a textual summary of an earlier chat session or a negotiated sub-protocol preserved for future interactions. Although individual nodes could store such state at the implementation level, this dimension focuses on whether the protocol itself promotes any form of persistence.

Its proposed values are \textit{none}, where no persistence features are offered; \textit{metadata}, where the protocol supports storage of metadata such as a negotiated schema; \textit{context}, where concrete interaction context such as session summaries is preserved; and \textit{hybrid}, where both metadata and context are explicitly maintained.

In practice, most protocols already implement \textit{metadata} persistence by exchanging component descriptions — such as the \textit{Agent Discovery Card} in A2A~\cite{a2adocsagentdiscovery} or the \textit{Agent Detail} endpoint in ACP~\cite{acpcoreconcepts}. Certain protocols additionally offer explicit features for context persistence: for example, LAP exposes a \textit{store} endpoint that manages a persistent key–value repository accessible across interaction \textit{threads}~\cite{lapdocsapi}.

Nonetheless, the dimension is rejected. Nearly every protocol already exchanges minimal metadata for node identification, and the protocol itself does not determine what connected nodes ultimately store persistently. Moreover, providing persistent state mechanisms is not an essential objective of these protocols. 


\subsection{Classifying Concrete Protocols}
\label{sec:taxonomy-design-classifying-protocols}

We now revisit the protocols used during taxonomy construction and assign them to the appropriate characteristics within the accepted dimensions. In the interest of conciseness, we only discuss the most salient classifications, those that best clarify and validate the taxonomy. A complete overview is provided in Table~\ref{tab:protocol-classification-markers} and Table~\ref{tab:protocol-classification-text}, the former optimized for pattern detection and the latter for informative depth.

\begin{sidewaystable}[p]
\centering
\scriptsize
\renewcommand{\arraystretch}{1.2} 
\setlength{\tabcolsep}{3pt}

\caption{Classification matrix of the nine analyzed protocols, for cross-protocol pattern comparison.}
\label{tab:protocol-classification-markers}

\begin{tabularx}{\textwidth}{@{} 
    >{\raggedright\arraybackslash}p{2.0cm} 
    >{\raggedright\arraybackslash}p{3.0cm} 
    *{9}{>{\centering\arraybackslash}X} @{}}
\toprule
 \textbf{Dimension} & \textbf{Characteristics} & 
 \textbf{MCP} \newline \footnotesize \cite{mcpgit,mcpdocscorearchitecture} & 
 \textbf{A2A} \newline \footnotesize \cite{a2agit,a2adocsmcp,a2adocskeyconcepts,a2adocsagentdiscovery,a2adocsspecification} & 
 \textbf{LAP} \newline \footnotesize \cite{lapgit,lapdocsapi} & 
 \textbf{agents} \newline \textbf{.json} \newline \footnotesize \cite{agentsjsongit,agentsjsonintro,agentsjsonschema} & 
 \textbf{Agora} \newline \footnotesize \cite{marro2024scalable,agoragit,agoraspecification} & 
 \textbf{ANP} \newline \footnotesize \cite{anpgit,anptechwhitepaper,anpcommunication,anpdiscovery} & 
 \textbf{LMOS} \newline \footnotesize \cite{lmosgit,lmosintroduction,lmosdiscovery} & 
 \textbf{ACP} \newline \footnotesize \cite{acpgit,acpcomparison,acpcoreconcepts} & 
 \textbf{agntcy} \newline \footnotesize \cite{agntcygit,agntcythreads,agntcydiscovery,agntcyconnectprotocol} 
 \\
\midrule
\multirow{3}{1.7cm}{\textit{Counterparty}} 
  & Agent   &   & x & x &   & x & x &   & x & x \\
  & Context & x &   &   & x &   &   &   &   &   \\
  & Hybrid  &   &   &   &   &   &   & x &   &   \\
\midrule

\multirow{2}{1.7cm}{\textit{Interaction State}} 
  & Stateless      & x &   &   & x &   &   &   &   &   \\
  & Session state  &   & x & x &   & x & x & x & x & x \\
\midrule
\multirow{5}{1.7cm}{\textit{Discovery Mechanism}} 
  & Static                   & x &   & x & x & x &   &   &   &   \\
  & Centralized              &   & x &   &   &   & x &   & x & x \\
  & Partially centralized  &   &   &   &   &   &   &   &   &   \\
  & Decentralized            &   &   &   &   &   &   &   &   &   \\
  & Hybrid                   &   &   &   &   &   &   & x &   &   \\
\midrule
\multirow{3}{1.7cm}{\textit{Payload}} 
  & Struct. data, artifacts & x &   &   & x &   &   &   &   &   \\
  & Conversation focused       &   &   &   &   &   &   &   &   &   \\
  & Hybrid                     &   & x & x &   & x & x & x & x & x \\
\midrule
\multirow{3}{1.7cm}{\textit{Schema Flexibility}} 
  & Single    &   &   & x & x &   &   &   &   &   \\ 
  & Multiple  & x & x &   &   &   &   & x & x & x \\ 
  & Evolving  &   &   &   &   & x & x &   &   &   \\ 
\bottomrule
\end{tabularx}
\end{sidewaystable}


\begin{sidewaystable}[p]
\centering
\scriptsize
\renewcommand{\arraystretch}{1.2}
\setlength{\tabcolsep}{3pt}

\caption{Per-protocol classification and rationale for the nine analyzed protocols.}
\label{tab:protocol-classification-text}

\begin{tabularx}{\textwidth}{
    >{\raggedright\arraybackslash}p{2.3cm}
    >{\raggedright\arraybackslash}p{2.1cm}
    >{\raggedright\arraybackslash}p{1.8cm}
    >{\raggedright\arraybackslash}p{2.1cm}
    >{\raggedright\arraybackslash}p{2.1cm}
    >{\raggedright\arraybackslash}p{1.8cm}
    >{\raggedright\arraybackslash}X
}
\toprule
\textbf{Protocol} &
\textbf{Counterparty} &
\textbf{Interaction State} &
\textbf{Discovery Mechanism} &
\textbf{Payload} &
\textbf{Schema Flexibility} &
\textbf{Rationale} \\
\midrule

\textbf{MCP} \newline \footnotesize \cite{mcpgit,mcpdocscorearchitecture} &
Context &
Stateless &
Static &
Structured data, artifacts &
Multiple &
Stateless core. Exposes tools; requires known endpoints. \\
\addlinespace

\textbf{A2A} \newline \footnotesize \cite{a2agit,a2adocsmcp,a2adocskeyconcepts,a2adocsagentdiscovery,a2adocsspecification} &
Agent &
Session state &
Centralized &
Hybrid &
Multiple &
Inter-agent tasks and streaming. Uses JSON card or registry. \\
\addlinespace

\textbf{LAP} \newline \footnotesize \cite{lapgit,lapdocsapi} &
Agent &
Session state &
Static &
Hybrid &
Single &
REST API with fixed endpoints for multi-turn sessions. \\
\addlinespace

\textbf{agents.json} \newline \footnotesize \cite{agentsjsongit,agentsjsonintro,agentsjsonschema} &
Context &
Stateless &
Static &
Structured data, artifacts &
Single &
Stateless OpenAPI JSON spec for LLM API discovery. \\
\addlinespace

\textbf{Agora} \newline \footnotesize \cite{marro2024scalable,agoragit,agoraspecification} &
Agent &
Session state &
Static &
Hybrid &
Evolving &
Decentralized. Plain-text schemas dynamically negotiated at runtime. \\
\addlinespace

\textbf{ANP} \newline \footnotesize \cite{anpgit,anptechwhitepaper,anpcommunication,anpdiscovery} &
Agent &
Session state &
Centralized &
Hybrid &
Evolving &
Stateful meta-layer dynamically negotiates structures and formats. \\
\addlinespace

\textbf{LMOS} \newline \footnotesize \cite{lmosgit,lmosintroduction,lmosdiscovery} &
Hybrid &
Session state &
Hybrid &
Hybrid &
Multiple &
Transport-agnostic. Standardizes metadata with hybrid discovery modes. \\
\addlinespace

\textbf{ACP} \newline \footnotesize \cite{acpgit,acpcomparison,acpcoreconcepts} &
Agent &
Session state &
Centralized &
Hybrid &
Multiple &
Multi-modal exchange. Agents register on centralized servers. \\
\addlinespace

\textbf{agntcy} \newline \footnotesize \cite{agntcygit,agntcythreads,agntcydiscovery,agntcyconnectprotocol} &
Agent &
Session state &
Centralized &
Hybrid &
Multiple &
OpenAPI extension for state/streams via central manifests. \\

\bottomrule
\end{tabularx}
\end{sidewaystable}

\subsubsection{Counterparty}

Various protocols are explicitly designed to facilitate agent-to-agent collaboration and thus address the counterparty type \textit{agent}. Agora~\cite{marro2024scalable}, ANP~\cite{anptechwhitepaper}, LAP~\cite{lapgit}, and agntcy~\cite{agntcyconnectprotocol} are classified as such. Moreover, A2A and ACP further emphasize that agents should use these protocols specifically for inter-agent communication, in contrast to protocols such as MCP, which are intended for connecting to context~\cite{a2adocsmcp,acpcomparison}. An \textit{MCP server} is defined as the counterparty addressed by the agent, which can expose resources (e.g. documents or data), tools, and prompts~\cite{mcpdocscorearchitecture}. 

\subsubsection{Payload}

MCP and agents.json are the only protocols assigned to the \textit{structured data and artifacts} payload type. MCP is designed for the exchange of structured requests and data responses between LLM-based agents and external tools, data sources, or services~\cite{mcpdocscorearchitecture}. Further, agents.json does not handle actual data exchange, its sole purpose is to provide agents with a structured description of an existing API~\cite{agentsjsonintro}. 

All remaining protocols are classified as \textit{hybrid}. They typically define a core unit of exchange capable of encapsulating both simple strings and arbitrary elements. The exchange is thus not limited to conversational text, purely structured or contextual information can be transmitted as well.

\subsubsection{Interaction State}

All protocols designed for agent-to-agent communication support multi-turn interactions through stateful units of work persisting across message exchanges. These units are named differently across protocols, like threads~\cite{agntcythreads,lapdocsapi}, tasks~\cite{a2adocskeyconcepts}, or in the case of Agora, realized through a simple \textit{multiround} flag that enables the message structure to maintain \textit{session state}~\cite{agoraspecification}.

In contrast, the core \textit{MCP} protocol and \textit{agents.json} define no explicit mechanism for session statefulness and are therefore \textit{stateless} by design~\cite{mcpdocscorearchitecture,agentsjsonintro,agentsjsonschema}.

\subsubsection{Discovery Mechanism}

First, MCP illustrates an important distinction: although agents can discover a server's capabilities at runtime, the \textit{discovery mechanism} dimension classifies MCP as \textit{static}, since the requesting agent must still know the server's base address a priori~\cite{mcpdocscorearchitecture}.

A2A, by contrast, supports three discovery strategies, all relying on a \textit{JSON Agent Card} as the agent's self-description. An agent may host its card at a standardized path on its own domain, share it through private channels, or publish and query cards via a centralized registry. Although static configuration is a primary discovery mechanism, the presence of a registry-based option places the protocol within the \textit{centralized} discovery class~\cite{a2adocsspecification}.

All other protocols fall into either the \textit{static} or \textit{centralized} category, with the exception of LMOS, which is classified as \textit{hybrid}~\cite{lmosdiscovery}.

\subsubsection{Schema Flexibility}

Most protocols either specify a \textit{single} concrete interface~\cite{lapdocsapi,agentsjsonschema} or allow \textit{multiple} distinct schemas to be defined in advance~\cite{mcpdocscorearchitecture,lmosintroduction,agntcyconnectprotocol}.

More interesting are the two protocols implementing \textit{evolving} schema functionality. In Agora, a core concept is the \textit{Protocol Document}, a plain-text description of everything an agent must understand to follow a given protocol. These documents are actively constructed and negotiated at runtime. Initial interactions occur entirely in natural language, but once a more efficient protocol has been established, subsequent exchanges can adopt a more structured format, possibly without any message parameters~\cite{marro2024scalable}. ANP similarly supports dynamic negotiation of payload formats and interaction structures, enabling agents to select or generate a schema at runtime~\cite{anpcommunication}.

\section{Discussion}\label{discussion}

We have developed a qualitative taxonomy to efficiently classify and analyze communication protocols for LLM agents. This hierarchical classification system helps to structure and thereby clarify an increasingly complex, rapidly growing field. Future protocols can be intuitively classified and compared at an abstract level, enabling researchers to identify overarching trends and anticipate potential innovations.

\subsection{Insights and Implications} \label{sec:insights}

Let us consider some important findings and implications that emerged during taxonomy development, backed by the classifications of all nine protocols (see Table~\ref{tab:protocol-classification-markers} and Table~\ref{tab:protocol-classification-text}).

If a protocol implements agent-to-agent communication (7 of 9), a \textit{session state} functionality is always available. This is expected, because contemporary LLM based agents rely on multi-turn interaction, which in turn requires state persistence. Moreover, none of the agent-to-agent protocols restrict themselves to \textit{conversation-focused} payloads (7 of 9). Instead, all are consistently \textit{hybrid}; while they can transmit textual messages, they also support purely structured payloads whenever appropriate. Furthermore, a clear trend towards schema flexibility emerges from the data. The majority of protocols permit multiple schema definitions (7 of 9), and two (2 of 9) additionally allow schemas to evolve during runtime. Schema negotiation is thereby regarded as a central component of agent-to-agent protocols, enabling LLM-based agents to fully leverage their capabilities. With respect to discovery mechanisms, only the LMOS protocol~\cite{lmosdiscovery} truly incorporates a decentralized, peer-to-peer approach. Most protocols instead rely on either a centralized registry (4 of 9) or static configuration (4 of 9). Should the Internet of Agents emerge as a dominant paradigm, decentralized approaches should be explored more and may prove increasingly important.

\begin{figure}[t]
  \centering
  \resizebox{\linewidth}{!}{%
    \begin{tikzpicture}
      \draw[band] (0,0)    rectangle (12,-1.4);
      \draw[band] (0,-1.7) rectangle (12,-3.1);
      \draw[band] (0,-3.4) rectangle (12,-4.8);
      \draw[band] (0,-5.1) rectangle (12,-6.5);
      \draw[band] (0,-6.8) rectangle (12,-8.2);
    
      \node[ltag]  at (0.30,-0.70) {L5};
      \node[lname] at (0.95,-0.44) {Deliberation};
      \node[lsub]  at (0.95,-0.96) {schema negotiation};
      \node[agentchip, minimum width=1.5cm] at (4.60,-0.70) {Agora};
      \node[agentchip, minimum width=1.4cm] at (6.35,-0.70) {ANP};
    
      \node[ltag]  at (0.30,-2.40) {L4};
      \node[lname] at (0.95,-2.14) {Interaction};
      \node[lsub]  at (0.95,-2.66) {tasks and streaming};
      \node[agentchip, minimum width=1.55cm] at (4.60,-2.40) {A2A};
      \node[agentchip, minimum width=1.55cm] at (6.40,-2.40) {LAP};
      \node[agentchip, minimum width=1.55cm] at (8.20,-2.40) {ACP};
      \node[agentchip, minimum width=1.55cm] at (10.00,-2.40) {agntcy};
    
      \node[ltag]  at (0.30,-4.10) {L3};
      \node[lname] at (0.95,-3.84) {Execution};
      \node[lsub]  at (0.95,-4.36) {structured tool calls};
      \node[ctxchip, minimum width=1.5cm] at (4.60,-4.10) {MCP};
    
      \node[ltag]  at (0.30,-5.80) {L2};
      \node[lname] at (0.95,-5.54) {Discovery};
      \node[lsub]  at (0.95,-6.06) {manifests, registries};
      \node[ctxchip, minimum width=2.0cm] at (4.60,-5.80) {agents.json};
      \node[hybchip, minimum width=1.5cm] at (6.85,-5.80) {LMOS};
    
      \node[ltag]  at (0.30,-7.50) {L1};
      \node[lname] at (0.95,-7.24) {Identity \& transport};
      \node[lsub]  at (0.95,-7.76) {auth, secure delivery};
      \node[subagent, minimum width=1.4cm] at (4.60,-7.50) {ANP};
      \node[subagent, minimum width=1.7cm] at (6.25,-7.50) {agntcy};
      \node[subhyb,   minimum width=1.6cm] at (8.20,-7.50) {LMOS};
    
      \node[legendsw, draw=agentline, fill=agentfill] at (0.30,-8.85) {};
      \node[legendtx] at (0.90,-8.85) {agent};
      \node[legendsw, draw=ctxline, fill=ctxfill] at (2.40,-8.85) {};
      \node[legendtx] at (3.00,-8.85) {context};
      \node[legendsw, draw=hybline, fill=hybfill] at (4.70,-8.85) {};
      \node[legendtx] at (5.30,-8.85) {hybrid};
      \node[legendsw, draw=hybline, dashed] at (6.90,-8.85) {};
      \node[legendtx] at (7.50,-8.85) {shared substrate};
    \end{tikzpicture}%
  }
  \caption{A candidate layered protocol stack for the Internet of Agents. Layers are ordered by the communication trilemma: lower layers favor efficiency and portability (rigid schemas, stateless, structured payloads), upper layers favor versatility (session-aware, schema-evolving). Each protocol is placed based on its primary role in Table~\ref{tab:protocol-classification-text}; chip color denotes the counterparty type, and dashed chips indicate protocols that additionally provide the identity/transport substrate. The layering is a proposed instantiation of the federated-stack argument in Section~\ref{sec:insights}, not an observed fact.}
  \label{fig:protocol-stack}
\end{figure}
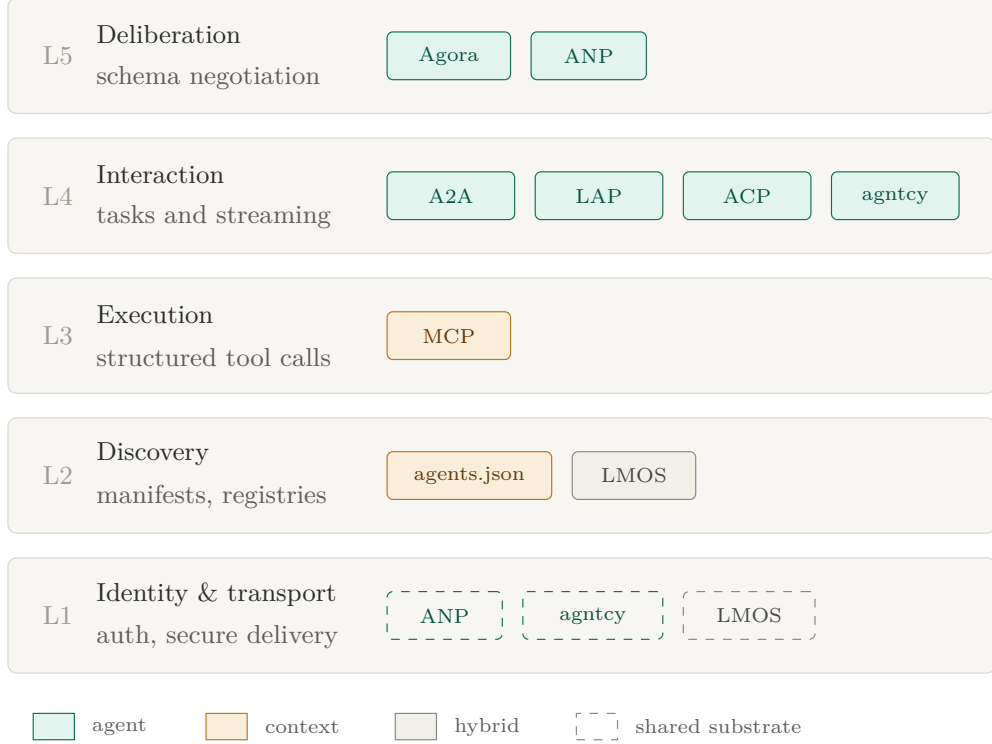

Looking forward, a central question is whether the field will converge on a single, monolithic standard or adopt a more modular architecture. Initially, the functional overlap between emerging protocols might suggest a short-term convergence toward a unified standard capable of supporting both agent-to-agent and agent-to-context interactions. Notably, A2A and agntcy present themselves as extensions of MCP~\cite{a2adocsmcp,agntcyintroduction}, though the primary organization behind MCP, Anthropic, has not officially endorsed either of them. 
While modifying an agent-to-agent protocol like A2A to incorporate MCP functionality would be relatively straightforward, drawing on historical precedents in computer networking, we argue that a monolithic~\enquote{winner-takes-all} standard is unlikely to prevail. Instead, mirroring the layered OSI model, illustrated in Figure~\ref{fig:protocol-stack}, our analysis suggests that the future of multi-agent system communication will evolve toward a federated, layered protocol stack. Under this architecture, the Internet of Agents will likely utilize lightweight specifications, such as agents.json, for static capability discovery; defer to highly structured context-protocols, e.g., MCP, for secure tool execution; and reserve session-aware, schema-evolving protocols, e.g., ANP or Agora, for complex, multi-turn deliberations.

A notable gap across all investigated protocols is the widespread absence of privacy safeguards, compliance checks, and policy enforcement mechanisms. This shortcoming will become more severe as agents are increasingly deployed in safety-critical domains such as healthcare or human resources, where mechanisms for protecting personal and security-sensitive data are essential~\cite{guo2024large}.

\subsection{Evaluating the Communication Trilemma}

Applying Marro's~\cite{marro2024scalable} Agent Communication Trilemma, introduced in Section~\ref{sec:related-work}, to our taxonomy reveals stark architectural trade-offs. The trilemma dictates that a protocol cannot simultaneously maximize versatility, efficiency, and portability. 

This trade-off is highly visible at the extremes of our \textit{Schema Flexibility} and \textit{Payload} dimensions. Protocols designed primarily for context-interaction, such as MCP, maximize portability and efficiency by enforcing rigid schemas, stateless interactions, and strictly structured data payloads. This strictness eliminates token-heavy negotiation, making them ideal for predictable, high-throughput agent-to-tool invocations. Conversely, protocols with \textit{evolving} schemas, such as Agora and ANP, maximize versatility to facilitate dynamic, open-ended multi-agent debates. However, this adaptability introduces significant token overhead and latency during schema negotiation, severely reducing efficiency.

Interestingly, the taxonomy reveals that the majority of surveyed protocols, such as A2A, LMOS, and ACP, navigate the center of this trilemma. By adopting \textit{hybrid} payload structures, \textit{session state} persistence, and \textit{multiple} pre-established schemas, they provide enough versatility to support complex interactions without incurring the extreme efficiency penalties of runtime protocol negotiation. Ultimately, mapping our taxonomy against the trilemma reinforces the conclusion from Section~\ref{sec:insights}: the impossibility of a single, omnipotent protocol strongly supports the necessity of a layered protocol stack for the future Internet of Agents.

\subsection{Limitations and Future Work} 

Future research could enrich our taxonomy with further dimensions that capture the domain in greater detail. Promising extensions include more technical protocol dimensions, authentication and security mechanisms, and the means by which protocols embed policies and norms. Also, classifying new protocols as they emerge would both strengthen the taxonomy's validity and allow systematic tracking and analysis of ongoing developments within the domain. Moreover, the taxonomy provides a practical basis for evaluating which protocols are most suitable for specific use cases and application domains.

\section{Conclusion}\label{conclusion}

In this paper, we have developed a comprehensive technical taxonomy of communication protocols for LLM-based agents. Motivated by the growing fragmentation of agent frameworks and protocol proposals, we examined how emerging protocols structure communication between LLM agents, other agents, tools, services, APIs, and external information systems. To develop the taxonomy, we followed a structured approach, defining the taxonomy's purpose, meta-characteristic, ending conditions, and iterative development process before applying the resulting framework to nine existing protocol implementations. By following a well-structured taxonomy development approach, we ensured methodological quality, producing a framework which allows efficient exploration and analysis of the diverse landscape of LLM agent communication protocols.

In addition, we identified several clear patterns by applying the taxonomy to nine concrete protocols. Most agent-to-agent protocols support multi-turn statefulness and the capacity to exchange both textual messages and structured data. Furthermore, schema negotiation is considered essential for efficient flexibility.
Most importantly, our analysis suggests a possible short-term convergence pressure toward protocols that unify agent-to-agent and agent-to-context communication. In the long term, rather than converging on a single monolithic standard, the field will likely evolve toward a federated, layered protocol stack, similar to the OSI model in traditional computer networking. Such an architecture would seamlessly integrate specialized protocols, ranging from lightweight capability discovery specifications to highly structured tool execution and schema-evolving deliberation mechanisms, in order to support the full spectrum of agent communication scenarios efficiently.
However, whether this leads to a single dominant standard or to a small set of interoperable protocol families remains an open question.

As LLMs continue to advance, LLM agents will grow more intelligent and capable of tackling real-world challenges. Likewise, LLM-powered multi-agent systems hold the potential to surpass the scalability constraints of single models. Consequently, the means by which agents communicate is a critical design element in building decentralized, flexible, and efficient agent networks, laying the groundwork for a future in which agents may operate partially or even fully autonomously in handling complex tasks.

\backmatter

\bibliography{bib/main}

\end{document}